\documentclass{aa}
\usepackage{graphics,latexsym}
\newcommand{\lya}{\hbox{Ly$\alpha$}}

\begin{document}

   \thesaurus{01     
              (11.01.2;  
               11.03.1;  
               11.05.2;  
               11.12.2;  
               12.03.3;  
               12.05.1)} 
   \title{Search for clusters at high redshift}

   \subtitle{I. Candidate Ly$\alpha$ emitters near 1138--262 at $z=2.2$}

   \author{J.D. Kurk \inst{1} \and H.J.A. R\"ottgering \inst{1} \and
           L. Pentericci \inst{2} \and G.K. Miley \inst{1} \and
	   W. van Breugel \inst{3} \and C.L. Carilli \inst{4} \and
           H. Ford \inst{5} \and T. Heckman \inst{5} \and 
           P. McCarthy \inst{6} \and A. Moorwood \inst{7}}

   \offprints{J.D. Kurk (kurk@strw.leidenuniv.nl)}

   \institute{Sterrewacht Leiden, P.O. Box 9513, 2300 RA, Leiden, 
              The Netherlands               
         \and
              Max-Planck-Institut f\"ur Astronomie, K\"onigstuhl 17,
              D-69117, Heidelberg, Germany
	 \and
	      Lawrence Livermore National Laboratory, 
	      P.O. Box 808, Livermore CA, 94550, USA
	 \and
	      NRAO, P.O. Box 0, Socorro NM, 87801, USA
	 \and
	      Dept. of Physics \& Astronomy, The Johns Hopkins University,
 	      3400 North Charles Street, Baltimore MD, 21218-2686, USA
	 \and
	      The Observatories of the Carnegie Institution of Washington,
              813 Santa Barbara Street, Pasadena CA, 91101, USA
	 \and
	      European Southern Observatory, Karl-Schwarzschild-Str. 2,
              D-85748, Garching bei M\"unchen, Germany
             }

   \date{Received March 13, 2000; accepted May 2, 2000}

   \maketitle

\begin{abstract}

Radio, optical and X-ray observations of the powerful radio galaxy PKS
1138--262 at $z=2.156$ have suggested that this galaxy is a massive
galaxy in the center of a forming cluster. We have imaged 1138--262
and the surrounding 38 square arcminute field with the Very Large
Telescope\footnote{Based on observations carried out at the European
Southern Observatory, Paranal, Chile, programme P63.O-0477(A).} in a
broad band and a narrow band encompassing the redshifted \lya\
emission. We detect 50 objects with rest equivalent width larger than
20 \AA\ and a luminous, highly extended \lya\ halo around 1138--262. If
the radio galaxy is at the center of a forming cluster, as observations
at other wavelengths suggest, these objects are candidate \lya\
emitting cluster galaxies.

\end{abstract}

%
\section{Introduction}

Observations of clusters at high redshift ($z > 2$) can directly
constrain cosmological models (e.g. Bahcall \& Fan \cite{bah98}), but
searches based on colours or narrow band emission have not established
the presence of massive clusters (Le F\`evre et al.\ \cite{fev96};
Pascarelle et al.\ \cite{pas96}; Keel et al.\ \cite{kee99}). There are
several indications (e.g. Pentericci et al.\ \cite{pen99}) that
powerful radio galaxies at high redshift (HzRGs) tend to be in the
center of forming clusters. The powerful radio galaxy PKS 1138--262 at
redshift 2.156 is a prime example of a forming brightest cluster
galaxy and has extensively been studied (e.g.\ Pentericci et al.\
\cite{pen97}). The arguments for 1138--262 being at the center of a
cluster include (a) the very clumpy morphology as observed by the HST
(Pentericci et al.\ \cite{pen98}), reminiscent of a massive merging
system; (b) the extremely distorted radio morphology and the
detection of the largest radio rotation measures (6200 rad m$^{-2}$)
in a sample of more than 70 HzRGs, indicating that 1138--262 is
surrounded by a hot, clumpy and dense magnetized medium (Carilli et
al.\ \cite{car97}; Pentericci et al. \cite{pen00}); (c) the
detection of X-ray emission around 1138--262 (Carilli et al.\ 1998),
indicating the presence of hot cluster gas, although a contribution to
the X-ray luminosity by the AGN cannot be precluded. For this reason,
we chose 1138--262 to carry out a pilot study with the VLT, to search
for direct evidence of clusters at high redshift. There are various
techniques for detecting high redshift companion galaxies. The colour
selection technique used to find Lyman Break Galaxies (LBGs) (Steidel
\& Hamilton \cite{ste92}) is not feasible at the redshift of
1138--262, since the Lyman limit falls at 2878 \AA, which is well
below the atmospheric cutoff. Therefore we have adopted the strategy
of narrow band imaging at the wavelength of the redshifted \lya\
line. This technique is capable of detecting galaxies at redshifts
similar to the radio galaxy redshift having strong \lya\ emission.

%
\section{Observation, data reduction and calibration}

Narrow and broad band imaging was carried out on April 12 and 13, 1999
with the 8.2m VLT Antu (UT1) using FORS in imaging mode. A narrow band
filter was used which has a central wavelength of 3814 \AA\ and a FWHM
of 65 \AA. For 1138--262 the emission of Ly$\alpha$ at 1216 \AA\ is
redshifted to 3838 \AA, which falls within the range of the narrow
band filter. The broad band filter was a Bessel B with central
wavelength of 4290 \AA\ and a FWHM of 880 \AA, which receives both
continuum and redshifted \lya\ line emission. The detector was a
Tektronix CCD with 2048$^2$ pixels and a scale of 0.2$''$ per
pixel. Eight separate 30 minutes exposures were taken in narrow band
and six 5 minutes exposures in B, shifted by $\sim$ 20$''$ with
respect to each other to minimize flat fielding problems and to
facilitate cosmic ray removal. The average seeing was 0.8$''$ and the
1$\sigma$ limiting AB magnitude per square arcsecond was 27.8 for the
narrow and 28.1 for the broad band image.

Image reduction was carried out using the IRAF reduction package. The
individual images were bias subtracted and flat fielded by twilight
flat fields for the narrow band and an average of the unregistered
science exposures for the broad band. The images were then registered
by shifting them in position by an amount determined from the location
of several stars on the CCD. The registered images were co-added and
cosmic rays removed. To improve the signal to noise and the
sensitivity to faint extended objects, the resulting images were
smoothed with a Gaussian function having a FWHM of 1$''$. The
spectrophotometric standard star GD108 (Oke \cite{oke90}) was used to
calibrate the fluxes in broad and narrow band. Calibration in broad
band is accurate to 0.1 magnitude, but absorption lines and a small
break in the part of the spectrum of GD108 which falls in the narrow
band inhibits equally accurate calibration for this filter. Instead,
it was assumed that the spectrum is flat in this wavelength range and
the narrow band is calibrated relative to the broad band with an
accuracy of 0.2 magnitude, as estimated from the spectrum of GD108. We
assume that the median equivalent width (EW) of a random sample of
objects is equal to zero. Since the median EW of our sample calibrated
by the standard star is equal to 0.5 $\times\ 10^{-19}$ erg s$^{-1}$
cm$^{-2}$ \AA$^{-1}$, we subtract this number from the narrow band
flux densities of the extracted objects to compute the EW. In this
way, we do not rely on the absolute flux calibration.

\begin{figure}
\resizebox{\hsize}{!}{\includegraphics{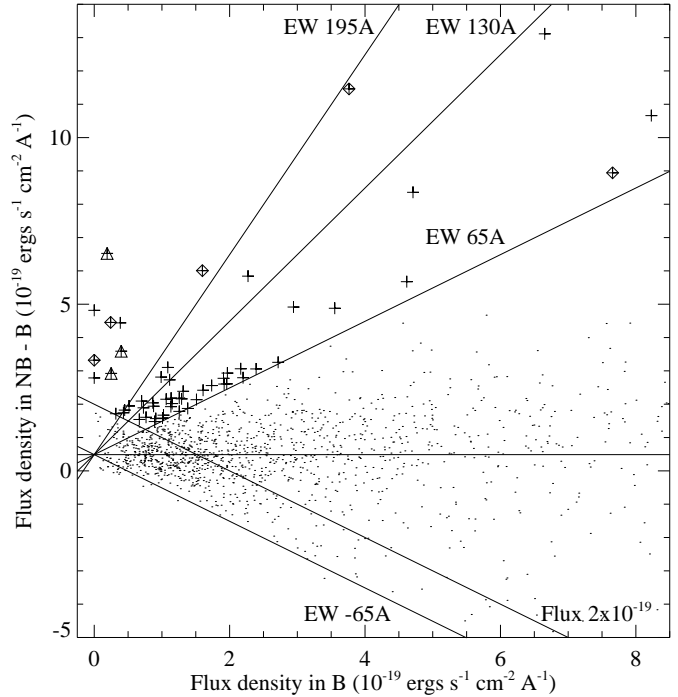}}
\caption{Flux densities of extracted objects are shown as dots. Plus
signs represent 55 objects selected as narrow band emitters, diamonds
are part of the \lya\ halo of 1138--262 (5 are not shown since they
have flux densities up to 8 $\times\ 10^{-18}$ erg s$^{-1}$ cm$^{-2}$
\AA$^{-1}$, exceeding the range of this plot) and triangles are the
examples shown in Fig.\ \ref{3cand}. Straight lines of EW 195, 130 and
65 \AA\ divide the region of candidates in strong (13), average(11)
and weak emitters (36). Also shown are lines of EW -65 \AA, narrow
band flux density 2 $\times\ 10^{-19}$ erg s$^{-1}$ cm$^{-2}$
\AA$^{-1}$ and the median NB -- B flux density of all extracted
objects.\vspace{-0.1cm}}

\label{selection}
\end{figure}

%
\section{Photometry and selection of \lya\ emitters}

Detection and photometry of objects in the field around 1138--262 was
carried out using SExtractor (Bertin \& Arnouts, 1996). We have taken
advantage of all observing time by normalizing and adding the narrow
and broad band images to obtain maximum signal to noise for the
majority of objects, which resulted in the extraction of 1727
sources. All detections have at least 9 connected pixels with a value
equal or larger than 2.5 times the rms sky noise. During this first
application of SExtractor, aperture sizes and shapes were determined
for each object separately. We gave the choice of aperture careful
consideration by comparing the signal-to-noise for fixed circular
apertures with a range of sizes and for elliptical apertures with a
range of scale factors. The half lengths of the principal axes of the
elliptical apertures are $\epsilon kr_1$ and $kr_1/\epsilon$, where
$\epsilon$ is the ellipticity determined by the isophots of the
object, $r_1$ the first moment of the radial (non-elliptical) light
distribution and $k$ a scale factor. The highest average
signal-to-noise for faint objects was obtained with elliptical
apertures using $k=1.25$. The apertures determined in this way were
used to carry out photometry on the narrow and broad band images.

\begin{figure}
\resizebox{\hsize}{!}{\includegraphics{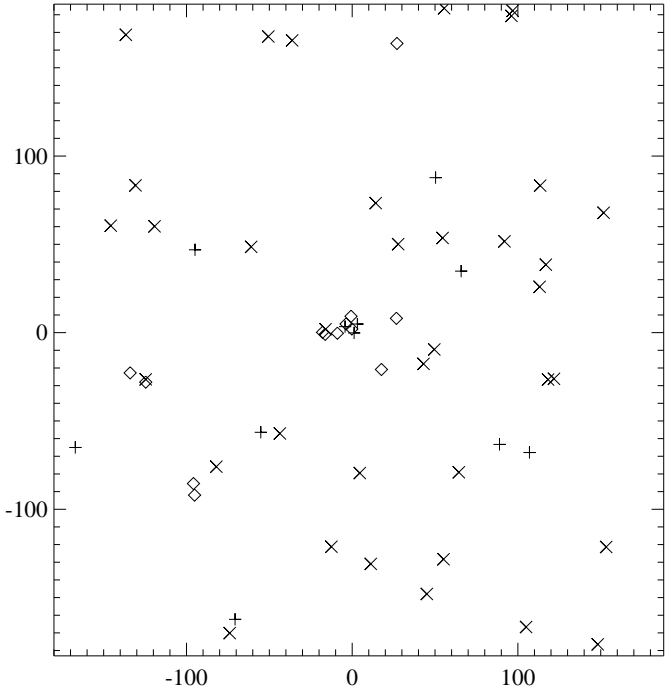}}
\caption{Relative positions in arcseconds of the 60 narrow band
emitters (including sources in the halo of 1138--262) in the final
registered image. The center of the image is at RA 11\fh40\fm47\fs9,
dec -26\fdg29\farcm08 (J2000). Diamonds, plus signs and crosses are
emitters with EW respectively higher than 195, 130 and 65 \AA. The
\lya\ halo of the radio galaxy is visible as a concentration of
symbols in the center of the image.\vspace{-0.1cm}}
\label{positions}
\end{figure}

\begin{figure}
\resizebox{\hsize}{!}{\includegraphics{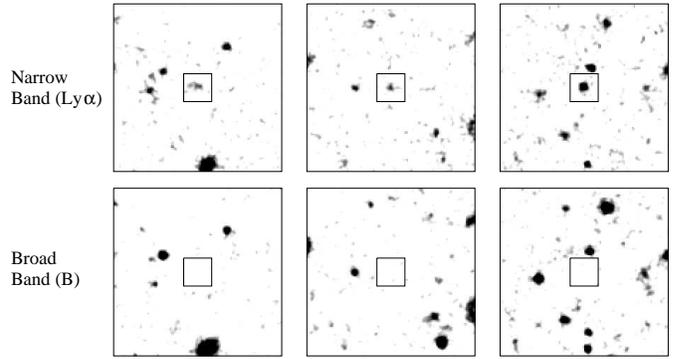}}
\caption{Three probable cluster companions shown at the top in narrow
band (Ly$\alpha$) and at the bottom in broad band (B).\vspace{-0.3cm}}
\label{3cand}
\end{figure}

Objects displaying twice as much flux density in the narrow band as in
broad band have EW 65 \AA. This equals to a rest frame EW of 20.5 \AA,
very close to the 20 \AA\ used as selection criterium by Steidel et
al.\ (\cite{ste99}, S99). There are 60 objects with EW larger than 65
\AA\ and narrow band flux density of at least 2 $\times\ 10^{-19}$ erg
s$^{-1}$ cm$^{-2}$ \AA$^{-1}$. The signal-to-noise of these objects is
at least 10 in narrow band, making the EW computation meaningful. We
consider these to be candidate \lya\ emitting galaxies. We subdivide
the candidates in very promising ones with EW three times this value
(13) and promising ones with EW two times this value (11), as shown in
Fig.\ \ref{selection}. Ten candidates coincide with peaks in the
extended ($\sim$ 160 kpc\footnotemark) halo of ionized hydrogen around
the radio galaxy. So in total, there are 50 previously unknown
emitters with a range of narrow band fluxes from 0.1 to 5$\times
10^{-16}$ ergs s$^{-1}$ cm$^{-2}$. The positions of the emitters are
shown in Fig.\ \ref{positions}. Images of three candidates are shown
in Fig.\ \ref{3cand}.

We believe that there will be few or no low redshift interlopers,
since [\ion{O}{ii}]$\lambda$3727 at redshift 0.02 is the only strong
line which falls in the narrow band filter, apart from \lya\ at
redshift 2.14. Galaxies at redshift 0.02 would easily be identified by
their large angular size. Since the narrow band is positioned on the
blue side of the broad band, our sample will not be contaminated by
objects with abnormal colors as Extremely Red Objects or red M dwarfs
or by objects with a break in their spectrum around 3900 \AA\ as LBGs
at $z \sim 3.3$. Contamination by bright quasars will be small, since
their surface density (0.04 arcmin$^{-2}$; Thompson et al.\
\cite{tom99}) is low enough that not more than 2 quasars will appear
in our field. Although there still is a possibility that we have
selected some rare, very blue objects with U - B $<$ -1.3, we shall
assume for the remainder of this paper that all objects with EW $> 65$
\AA\ are \lya\ emitting galaxies.

\footnotetext{Throughout this article, we adopt a Hubble constant
of $H_0$=50 km s$^{-1}$Mpc$^{-1}$ and a deceleration parameter of
$q_0$=0.5.}

%
\section{Clustering analysis}

Is there an enhanced density of galaxies around 1138--262? Do the
candidates make up a forming cluster?

The number of 50 \lya\ emitters in the 35.4 arcmin$^2$ field ($\sim
2.3$ arcmin$^2$ is rendered unusable by bright stars) is equivalent to
a surface density of 1.4 per arcmin$^2$. The comoving volume for this
field size (8.3 Mpc$^2$) and redshift range ($2.110 \le z \le 2.164$)
is 4758 Mpc$^3$, resulting in a volume density of $1.1 \times 10^{-2}$
emitters per Mpc$^3$. A contour plot of surface density does not show
a strong concentration of emitters centered on the radio galaxy or
anywhere else in the field.

The angular two point correlation function was determined by the
Landy-Szalay estimator (Landy \& Szalay \cite{lan93}). Due to the
occurence of close pairs in our sample, the correlation function shows
a signal at short distances, but there is no evidence that the
distribution of emitters is significantly different from a random
distribution on scales larger than 20\arcsec.

\begin{figure}
\resizebox{\hsize}{!}{\includegraphics{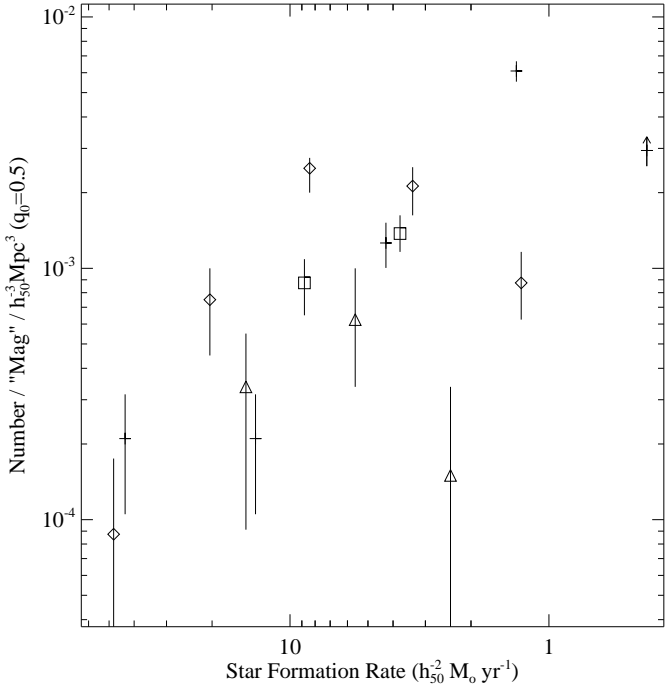}}
\caption{The luminosity function of our \lya\ emission selected
objects at $z=2.2$ ($+$), Cowie \& Hu's (\cite{cow98}) blank field
narrow band emitters at $z=3.4$ ($ \triangle$), Campos et al.'s
(\cite{cam99}) near-QSO search at $z=2.6$ ($\Box$) and S99's
overdensity of star forming galaxies at $z=3.1$ ($\Diamond$). Note
that our bin with the lowest SFR is only a lower limit, due to
sensitivity incompleteness. Figure reproduced from S99.}
\label{lumfun}
\end{figure}

S99 present an overview of luminosity functions of three \lya\
emission searches in their Fig.\ 5. We have converted our \lya\
luminosities to star formation rate (SFR) using the relation
L$_\nu$(1500\AA) = 8.0$\times 10^{27}$ (SFR / M$_\odot$ yr$^{-1}$)
ergs s$^{-1}$ Hz$^{-1}$ (Madau et al.\ \cite{mad98})
and computed the number of emitters (omitting the radio galaxy halo
objects) per SFR bin per Mpc$^3$ assuming $z=2.14$ for all
emitters. The central wavelength of the B filter is 1366 \AA\ in the
restframe, which is close enough to 1500 \AA\ to omit a K-correction
carried out with an uncertain spectral index. Fig.\ \ref{lumfun} shows
that the density of the cluster companions of 1138--262 (plus signs
with vertical error bars) is comparable to the near-QSO search of
Campos et al.\ (\cite{cam99}) at $z=2.6$, which represents a small
overdensity relative to the blank field search of Cowie \& Hu
(\cite{cow98}) at $z=3.4$. The density is, however, lower than the
Lyman break galaxy \emph{spike} of S99 at $z=3.1$. We have assumed
that properties of forming clusters and galaxies do not change too
fast to allow meaningful comparison between clusters at different
epochs. The figure also shows that we detect SFR as low as 1 h$^{-2}$
M$_\odot$ yr$^{-1}$, while Cowie \& Hu (\cite{cow98}) and S99 reach
their sensitivity limit at about 8 h$^{-2}$ M$_\odot$ yr$^{-1}$. This
is due, in part, to our sampling of a lower redshift range.

\section{Discussion}
Although we have detected a number of \lya\ emitters in the field of
1138--262, from the present data it is impossible to unambiguously
determine whether a significant fraction of these objects form part of
the presumed cluster around 1138--262. Note that the observations
described here can only detect galaxies with bright \lya\ emission
lines not attenuated by dust and these might comprise merely a small
fraction of the galaxy content of the presumed cluster. First, of the
whole population of LBGs as studied by S99 only 20\% have a Lya rest
EW 20 \AA. About 50\% of the LBGs have the Lya region even in
absorption. Second, only a fraction of the cluster galaxies are likely
to be actively forming stars. We assume that the ratio of early to
late type galaxies in the cluster around 1138--262 is the same as the
1.1:1 ratio in intermediate redshift ($z\sim0.5$) clusters as found by
Andreon (\cite{and98}). Taking these two factors into account, the
detected \lya\ galaxies might comprise only 10\% of the cluster
galaxies present around 1138--262. In addition to this, we miss
galaxies emitting \lya\ outside the limited range of wavelength
covered by our narrow band filter. Assuming that the radio galaxy is
at rest with respect to the cluster, the wavelength of \lya\ from
galaxies with a positive velocity greater than 860 km s$^{-1}$ falls
outside the FWHM of the filter. Since present day clusters have
velocity dispersions as large as $\sim 1200$ km s$^{-1}$ (Mazure et
al.\ \cite{maz96}), we miss 15\% of the velocity range of the presumed
cluster around 1138--262 if we assume the same velocity dispersion for
clusters at high redshift.

%
\section{Conclusions}

We have detected 50 candidate \lya\ emitters close to radio galaxy
1138--262 at redshift 2.2. These \lya\ emitters are candidate
starburst galaxies in the cluster, which is presumed to exist or to be
forming around 1138--262. We do not find a significant overdensity of
candidates compared to luminosity functions of blank fields nor do we
detect a strong concentration gradient in our 8 Mpc$^2$ field. The
next step in our search for clusters at high redshift is to confirm
the existence of the \lya\ emitters by multi object spectroscopy at
the VLT and determine the spatial correlation function and velocity
dispersion, which together with the size of the cluster will give a
direct estimate of the total mass. Additionally, we will carry out
X-ray observations of 1138--262 with the Chandra telescope.

\begin{acknowledgements}
   JK acknowledges productive discussions with Philip Best and Tom
   Thomas. The work by WvB at IGPP/LLNL was performed under the
   auspices of the US Department of Energy under contract
   W-7405-ENG-48. IRAF is distributed by the National Optical
   Astronomy Observatories, which are operated by the Association of
   Universities for Research in Astronomy, Inc., under cooperative
   agreement with the National Science Foundation.
\end{acknowledgements}

\end{document}